\documentclass[aps,nofootinbib,showpacs]{revtex4}

\usepackage{graphicx}
\usepackage{dcolumn}
\usepackage{bm}

\makeatletter
\def\simleq{\mathrel{\mathpalette\gl@align<}}
\def\simgeq{\mathrel{\mathpalette\gl@align>}}
\def\gl@align#1#2{\lower.6ex\vbox{\baselineskip\z@skip\lineskip\z@
\ialign{$\m@th#1\hfill##\hfil$\crcr#2\crcr\sim\crcr}}}
\makeatother

\newcommand{\fslash}[1]{\ooalign{\hfil/\hfil\crcr$#1$}}

\newcommand{\bra}{\langle}
\newcommand{\ket}{\rangle}
\newcommand{\braket}[1]{\bra #1 \ket}

\newcommand{\qq}{\braket{\bar{q}q}}

\newcommand{\sbs}{\braket{\bar{s}s}}

\newcommand{\sGs}{\braket{\bar{s}g_s \sigma\cdot G s}}
\newcommand{\GG}{\braket{\frac{\alpha_s}{\pi} G^2}}

\newcommand{\fsl}[1]{\fslash{#1}}

\def\Eq.#1{Eq.~(\ref{#1})}

\renewcommand{\sb}{\bar s} 

\newcommand{\MB}{M}

\newcommand{\half}{{1\over 2}}

\newcommand{\Tp}{\Theta^+}
\newcommand{\sth}{\sqrt{s_{\rm th}}}
\newcommand{\sthd}{s_{\rm th}}

\begin{document}


\title{Penta-quark baryon from the QCD Sum Rule}

\author{Jun Sugiyama}%
\email[email: ]{sugiyama@th.phys.titech.ac.jp}%
\affiliation{%
Department of Physics, Tokyo Institute of Technology,
Meguro, Tokyo, 152-8551, Japan
}%

\author{Takumi Doi}%
\email[email: ]{doi@th.phys.titech.ac.jp}%
\affiliation{%
Department of Physics, Tokyo Institute of Technology,
Meguro, Tokyo, 152-8551, Japan
}%

\author{Makoto Oka}%
\email[email: ]{oka@th.phys.titech.ac.jp}%
\affiliation{%
Department of Physics, Tokyo Institute of Technology,
Meguro, Tokyo, 152-8551, Japan
}%


\begin{abstract}
Exotic penta-quark baryon with strangeness  $+1$, $\Tp$, is studied in the QCD sum rule approach.
We derive sum rules for the positive and negative parity baryon states with $J=\half$ and $I=0$.
It is found that the standard values of the QCD condensates predict a negative parity $\Tp$ of mass
$\simeq 1.5$ GeV, while no positive parity state is found.
We stress the roles of chiral-odd condensates in determining the parity and mass of $\Tp$.

\end{abstract}



\maketitle


The discovery of the $\Tp$ state by the LEPS group at SPring-8\cite{Nakano} is quite striking.
$\Tp$ is produced by the $\gamma + n \to K^- + \Theta^+$ reaction and is observed 
in the invariant mass of the $n + K^+$ final state.  Its mass is 1540 MeV/c$^2$ and the width is 
less than 25 MeV.  Several other groups have confirmed this result.\cite{DIANA,CLAS,SAPHIR} 
The conservation laws of the strong interaction tell us that $\Tp$ is a baryon
with strangeness $+1$ and thus contains a $\sb$ quark.  
Therefore the simplest quark content of $\Tp$ is $uudd\sb$, and it cannot be made of three quarks.  
Over a thousand hadrons
are compiled by the Particle Data Group,\cite{PDG} but so far none of them is confirmed as an exotic 
hadron, which cannot be associated with either a three quark baryon or a quark-antiquark meson.

The spin and parity of the $\Tp$ are not yet determined, but many conjectures have been made.%
\cite{Diakonov,Lipkin,Jaffe,Hosaka,Riska,Capstick,Carlson}
We first realize that $\Tp$ may be isospin singlet ($I=0$), because no $pK^+$ resonance is observed.  
This is against the proposal by Capstick et al.\cite{Capstick}, who interpreted $\Tp$ 
as an isotensor ($I=2$) state because of the ``unusually narrow'' width.  
It is, however, pointed out that the coupling constant for the
$\Tp \to NK$ decay is not too small even if $\Tp$ is an $I=0$ baryon.\cite{hosaka_p}
Thus we assume that $\Tp$ is an $I=0$ baryon.

The spin is naturally assumed to be $\half$, because all the hadrons observed so far follow 
the simple rule that higher spin states have larger masses.%
\footnote{We assume that there is no other resonance state nor a bound state below $\Tp(1540)$.  Such 
assumption is supported by the current experimental data.}
The one-gluon exchange interaction,
which is a typical $qq$ interaction, prefers lower spin states, for instance.
This, however, should be confirmed by experiment.
It is also interesting to note the spin of the four-quark subsystem.  If the $(ud)^2$ system
has $J=1${\cite{Hosaka}}, then $\Tp$ should have a $J=3/2$ partner at maybe a few hundred
MeV above.  No excited state of $\Tp$ is observed so far.

The narrowness of $\Tp$ may indicate a $P$-wave resonance, meaning $\half^+$ 
state.  This is consistent with the Skyrme model prediction by Diakonov et al.\cite{Diakonov}
However, the quark model naturally gives $\half^-$ state as the ground state.\cite{Carlson} 
Several suggestions were made\cite{Lipkin,Jaffe,Riska} to reverse their order, but no quantitative microscopic evaluation has been made.

It is rather clear that the parity as well as the spin of the $\Tp$ is critical in
understanding the penta-quark structure of this baryon.
We here attempt to determine the parity, assuming that its spin is $\half$, directly from 
quantum chromodynamics (QCD).
To this goal, we employ the QCD sum rule technique.\cite{SVZ,RRY}
In this approach, a correlation function is calculated by the use of operator product expansion (OPE) in the deeply Euclidean region on one hand, and is compared with that calculated for a phenomenological parameterization.
Thus the sum rules relate hadron properties directly to the QCD vacuum condensates, such as $\qq$ and $\GG$, as well as the other fundamental constants, such as $m_s$.

We employ the following interpolating field operator for the penta-quark state,
\begin{eqnarray}
  \eta(x)&=&\epsilon^{abc}\epsilon^{def}\epsilon^{cfg}
   \{u_a^T(x)Cd_b(x)\}\{u_d^T(x)C\gamma_5 d_e(x)\}C\bar{s}_g^T(x) \nonumber \\
  \bar{\eta}(x)&=&-\epsilon^{abc}\epsilon^{def}\epsilon^{cfg}
   s_g^T(x)C\{\bar{d}_e(x)\gamma_5 C \bar{u}_d^T(x)\}\{\bar{d}_b(x)C \bar{u}_a^T(x)\} ,
   \label{eq:IF}
\end{eqnarray}
where $a,b,c,\cdots$ are color indices and $C=i\gamma^2\gamma^0$.
It is easy to confirm that this operator produces a baryon with $J=\half$, $I=0$ and strangeness $+1$.
The parts,
$S^c(x) = \epsilon^{abc} u_a^T(x)C\gamma_5 d_b(x)$ and
$P^c(x) = \epsilon^{abc} u_a^T(x)Cd_b(x)$,
give the scalar $S$ ($0^+$) and the pseudoscalar $P$ ($0^-$) $ud$ diquarks, respectively.  They both belong to the anti-triplet (3*) representation of the color SU(3) and have $I=0$.  The scalar diquark corresponds to the $^1S_0$ state of the $I=0$ $ud$ quark system.  It is known that a gluon exchange force as well as the instanton mediated force commonly used in the quark model spectroscopy give significant attraction between the quarks in this channel. The pseudoscalar diquark does not have nonrelativistic limit, though from the quantum number, we may assign it to the $^3P_0$ state of $ud$ ($I=0$).  
 
 It is natural to ask why we do not use a product of two {\it scalar\/} diquarks. The answer is that it is not possible to construct a local operator with two $ud$ scalar diquarks, as they behave as identical boson operators, which are to be antisymmetric in the color quantum number.  Thus, as a next simple local operator, we employ the combination of a scalar diquark and a pseudoscalar diquark.  
One of the advantages of this operator is that its coupling to the main continuum state, $NK$, is expected to be small, because $\eta(x)$ cannot be decomposed into a product of $N(3q)$ and $K(q\bar q)$ operators in the nonrelativistic limit.

We would like to stress here that the parity of the baryon is not specified because the interpolating field operator, Eq.~(\ref{eq:IF}), may generate both the positive and negative parity baryons.  To digest this fact, we consider the spatial inversion applied to $\eta(t,\vec x)$,
\begin{equation}
\eta(t, \vec x) \to + \gamma^0 \eta(t, -\vec x) .
\end{equation}
It may seem that the parity of $\eta$ is positive and therefore $\eta$ annihilates positive parity baryons only.
But one may change the parity of the operator simply by multiplying $\gamma^5$,
\begin{equation}
 \gamma^5 \eta(t, \vec x) \to + \gamma^5 \gamma^0 \eta(t, -\vec x) =
 - \gamma^0 \gamma^5 \eta(t, -\vec x) .
\end{equation}
The correlation function
\begin{equation}
 \Pi_T (q) =  \int d^4 x\,  e^{iq\cdot x} i \bra 0| T(\eta(x) \bar \eta(0) )|0\ket
 \label{eq:PiT}
\end{equation}
can be expressed in terms of the spectral function by inserting intermediate baryon states 
in between $\eta$ and $\bar \eta$.  For the positive parity states, the matrix element is given by
\begin{equation}
 \bra 0| \eta(x) |B^+(\vec p) \ket = \lambda_+ \, u_+(\vec p) e^{-ip\cdot x} ,
  \label{eq:lam1}
\end{equation}
while for the negative parity states, we have
\begin{equation}
 \bra 0| \eta(x) |B^-(\vec p) \ket = \lambda_- \, \gamma^5 u_-(\vec p) e^{-ip\cdot x} .
  \label{eq:lam2}
\end{equation}
Thus the correlation function can be expressed by the positive-parity spectral function $\rho^+$ and the negative-parity one $\rho^-$ as
\begin{eqnarray}
 \Pi_T(q) &= & - \int dm_+ \, {\rho^+(m_+)\over \fsl{q}-m_+} 
 + \int dm_- \, \gamma^5 {\rho^-(m_-)\over \fsl{q}-m_-} \gamma^5 \nonumber\\
        &= & - \int dm_+ \, {\rho^+(m_+)\over \fsl{q}-m_+} 
 - \int dm_- \,  {\rho^-(m_-)\over \fsl{q} + m_-} .
 \label{eq:spectral}
\end{eqnarray}

In order to separate the positive and negative parity states out of the correlation function, we use the technique developed in ref.~\cite{Jido-Kodama} for the ordinary three-quark baryons.  We consider the retarded Green's function and choose the rest frame, $\vec q=0$, 
\begin{equation}
 \Pi (q_0) =  \int d^4 x\,  e^{iq\cdot x} i \bra 0| \theta(x^0) \,\eta(x) \bar \eta(0) |0\ket|_{\vec q=0} \, .
 \label{eq:Pi}
\end{equation}
This correlation function is analytic for ${\rm Im}\, q_0 >0$ and satisfies 
\begin{equation}
 {\rm Im}\, \Pi (t, \vec x) =  {\rm Im}\, \Pi_T (t, \vec x)   \quad \hbox{for $t>0$}, 
 \label{eq:imaginary}
\end{equation}
where $\Pi_T$ is the Feynman correlator defined in Eq.~(\ref{eq:PiT}).
Thus the retarded correlation function has singularities only at real positive $q_0$.
From Eqs.~(\ref{eq:spectral}) and (\ref{eq:imaginary}), we obtain, for real $q_0>0$,
\begin{eqnarray}
\label{eq:AB}
 {1\over \pi}{\rm Im} \Pi(q_0) &= & A(q_0) \gamma^0 + B(q_0) \\
 A(q_0) &=&  {1\over 2} \left(\rho^+(q_0) + \rho^-(q_0)  \right)\nonumber\\
 B(q_0) &=&  {1\over 2} \left(\rho^+(q_0) - \rho^-(q_0)  \right)  ,\nonumber
\end{eqnarray}
or equivalently,
\begin{eqnarray}
 \rho^{\pm}(q_0)  &=& A(q_0) \pm B(q_0) .
 \label{eq:rhopm}
\end{eqnarray}

\begin{figure}[hbtp]
 \caption{
Contributions to Eq.~(\ref{eq:OPE}).  The dashed lines are gluons, and the blob on the quark line indicates the insertion of the mass, and the condensates. The figure (a) gives the term without condensate, (b) the $m_s$,
$\sbs$, $\sGs$, $m_s\sbs$, and $m_s\sGs$ terms, and (c) and (d) give the gluon condensate term, $\GG$.%
}
\begin{center}
\includegraphics*[width=5.5cm]{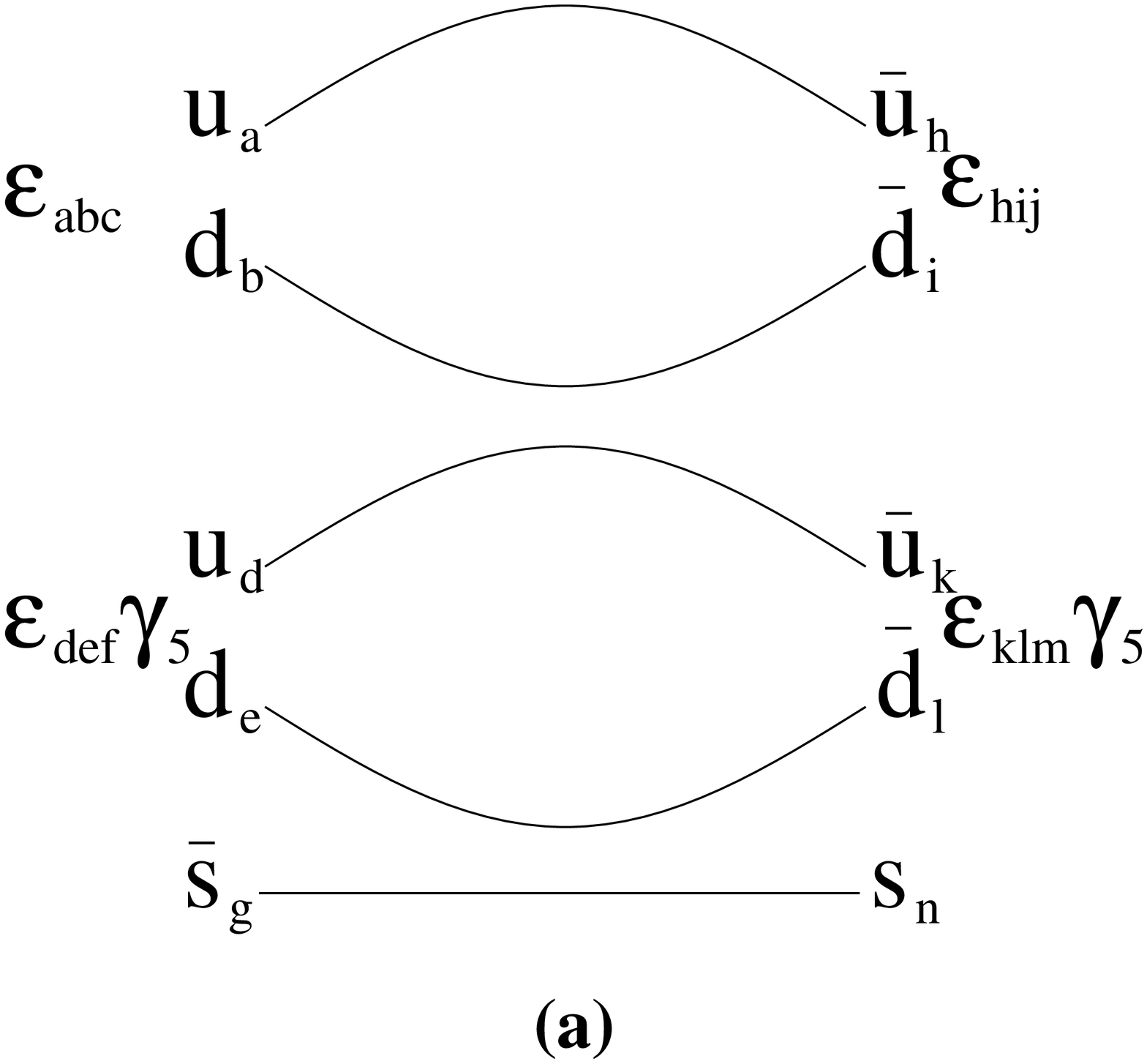}
\includegraphics*[width=5.5cm]{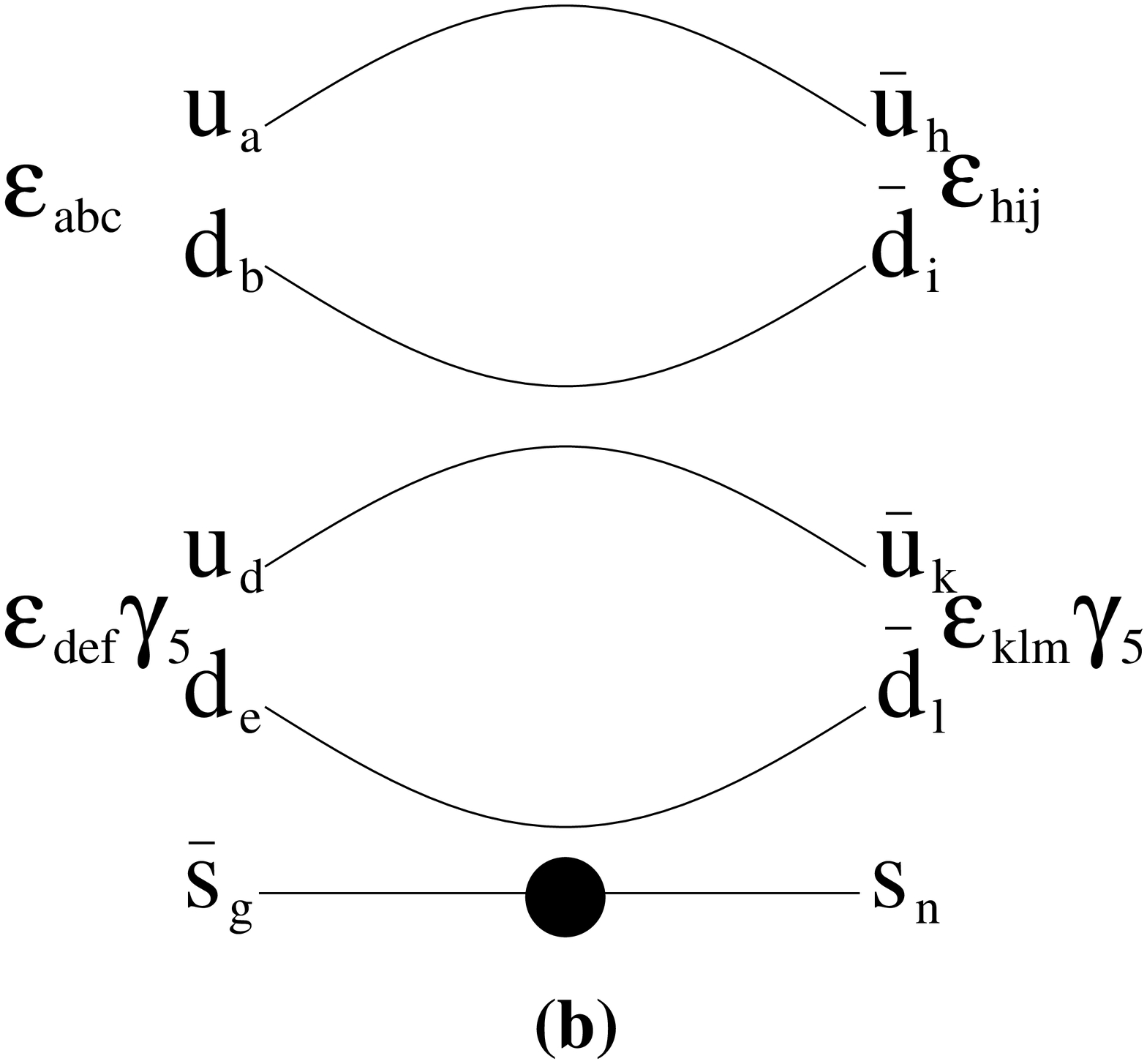}
\vskip 1cm
\includegraphics*[width=5.5cm]{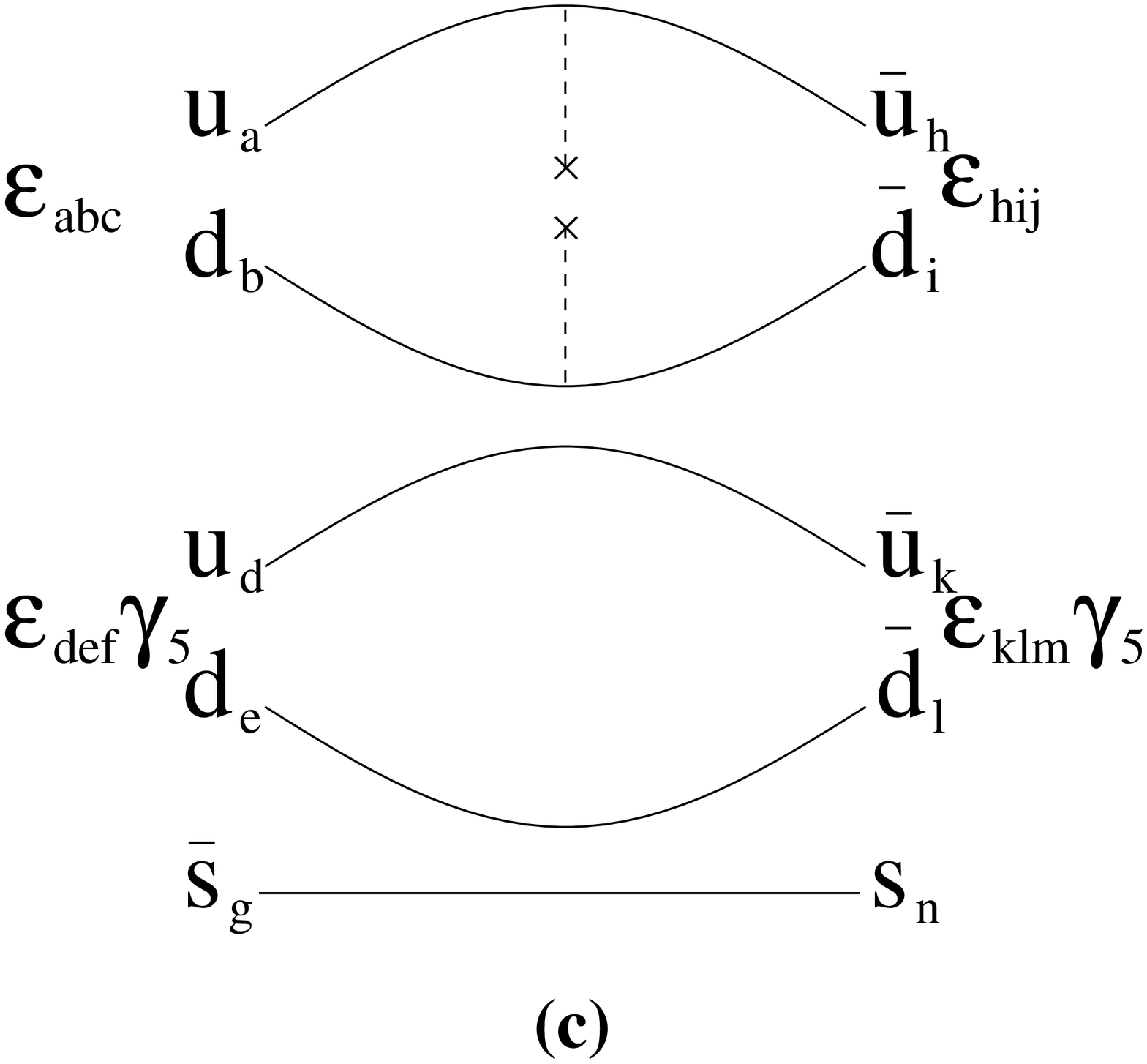}
\includegraphics*[width=5.5cm]{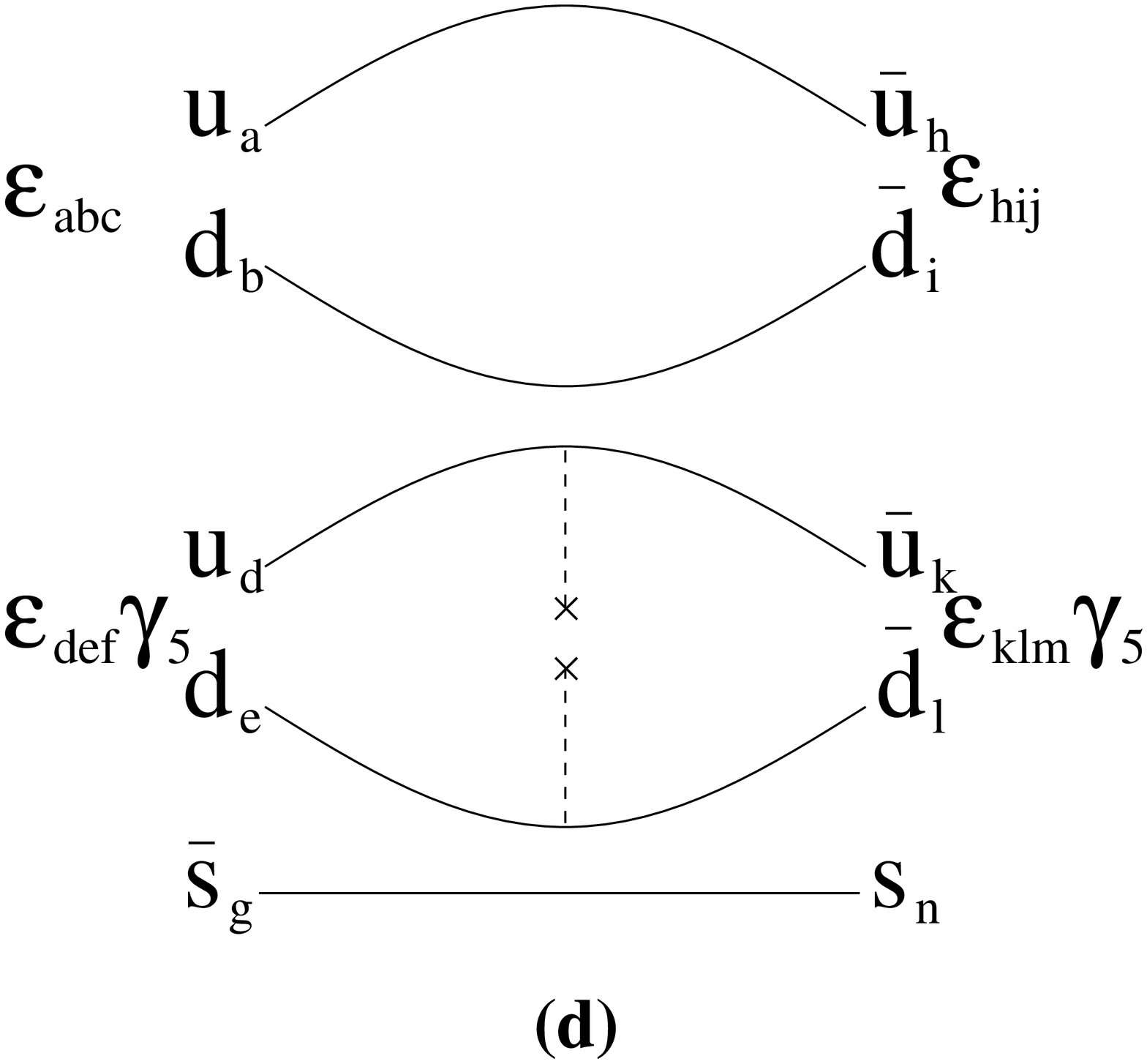}
\end{center}
\label{fig:OPE}
\end{figure}

The imaginary part of the correlation function is evaluated at the asymptotic region,
$q_0^2 \to - \infty$, by the operator product expansion (OPE) technique.  
We then obtain
\begin{eqnarray}
 A_{\rm OPE} (q_0) &=&  \frac{q_0^{11}}{5!\>5!\>2^{10} 7 \pi^8}
        + \frac{q_0^7}{3!\>5!\>2^8\pi^6}m_s \sbs 
        +\frac{q_0^7}{5!\>3!\>2^{10}\pi^6}\GG 
		- \frac{q_0^5}{4!\>3!\>2^9\pi^6} m_s \sGs
		\nonumber\\
 B_{\rm OPE}(q_0) &=&
        \frac{q_0^{10}m_s}{5!\>5!\>2^{10}\pi^8}
        - \frac{q_0^8}{4!\>5!\>2^{7}\pi^6}\sbs
        + \frac{q_0^6}{3!\>4!\>2^9\pi^6} \sGs 
		\label{eq:OPE}
\end{eqnarray}
from the OPE up to the dimension 6 operators.  
Fig.~\ref{fig:OPE} shows various terms of OPE graphically.
Here the masses of the $u$, $d$ quarks are neglected.
Some special features of this OPE are (1) that neither $u$ nor $d$ quark condensate appears 
up to this order, and (2) the $B$ term consists of chiral odd condensates, $\sbs$ and $\sGs$,
as well as the strange quark mass $m_s$, which breaks chiral symmetry explicitly.
Eq.~(\ref{eq:rhopm}) shows that the splitting of the positive and negative parity spectrum 
comes from $B$.  In other words, chiral symmetry breaking is responsible for the parity splitting.
This feature has been seen also in the baryon sum rule and 
shows explicit roles of the chiral symmetry breaking on the hadron spectrum.

The first feature comes from the structure of the interpolating field operator, Eq.~(\ref{eq:IF}).
One sees in Fig.~\ref{fig:OPE} that the OPE consists only of the contractions of the scalar diquarks, $S-S$, 
and the pseudoscalar diquarks, $P-P$, while the other terms of the type $S-P$ vanish.
Then the chiral structure of the diquark operators prohibits appearance of the $u$, or $d$, condensates,
i.e., the diquarks contain only the left-left or right-right combinations and therefore a single quark 
condensate vanishes in the chiral limit, $m_u\simeq m_d\simeq 0$.
We also note that four quark condensates of dimension 6 do not contribute to the leading order in
$1/N_c$, which is another advantage of this choice of $\eta(x)$.

The sum rule is obtained by comparing the OPE of the correlation function, Eq.~(\ref{eq:OPE}), and 
explicit forms of the spectral functions using the analytic continuation.
The spectral function is commonly parametrized by a pole plus continuum contribution,
\begin{eqnarray}
    \rho^{\pm}_{\rm Phen}(q_0) &=&  {|\lambda_{\pm}|^2 } \delta(q_0 - m_{\pm}) 
                 +\theta(q_0- \sth)  \rho^{\pm}_{\rm CONT}(q_0),
\label{eq:rho}
\end{eqnarray}
where $|\lambda_{\pm}|^2$, defined in Eqs.~(\ref{eq:lam1}) and (\ref{eq:lam2}), denotes the residue of the pole determined by the matrix element of 
the interpolating field for the designated state of mass $m_{\pm}$.  The residue should be positive,
which gives a condition to check the validity of the sum rule.  The continuum part is assumed to be
identical to the corresponding OPE function at above the threshold $\sth$,
$\rho^{\pm}_{\rm CONT} = \rho^{\pm}_{\rm OPE} \equiv A_{\rm OPE} \pm  B_{\rm OPE}$.

In order to enhance the pole part and also suppress the higher dimension terms of the OPE,
we introduce a weight function 
\begin{equation}
W(q_0) = \exp \left( - {q_0^2\over \MB^2} \right)  ,
\end{equation}
where $\MB$ is the relevant mass scale for the baryon.  
The sum rule is obtained as
\begin{equation}
\int dq_0 W(q_0) \rho^{\pm}_{\rm Phen} (q_0)= \int dq_0 W(q_0) \rho^{\pm}_{\rm OPE} (q_0) .
\end{equation}
This form of the weight function is borrowed from the Borel
sum rule formulation and there $\MB$ is often called the ``Borel mass''.  Physical quantities are to be
independent of the choice of $\MB$ ideally, but in practice, the truncation in the OPE and 
the incompleteness of the pole plus continuum assumption lead to mild $\MB$ dependence.
We have to choose a reasonable range of $\MB$ to evaluate the physical quantities.

Finally, we obtain the sum rules for the positive and negative parity baryons,
\begin{eqnarray}
  |\lambda_\pm|^2 e^{-\frac{{m_{\pm}}^2}{M^2}}&=&
           \frac{1}{3!4!2^7\pi^6}
		   \left[\frac{1}{5600\pi^2}\, I_{11}(M, \sthd)
		   \pm\frac{1}{800\pi^2}\, I_{10}(M, \sthd) m_s\right.     
		   \nonumber\\
       & &\hspace{3em} \mp\frac{1}{20}\, I_8(M, \sthd)\sbs
	   + \frac{1}{10}\, I_7(M, \sthd)m_s \sbs 
	   \nonumber\\
	   & &\hspace{5em}
		+\frac{1}{40}\, I_7(M, \sthd) \GG
          \pm\frac{1}{4}\, I_6(M, \sthd) \sGs 
		  \nonumber \\
        & &\left.\hspace{5em}
		- \frac{1}{4}\, I_5(M, \sthd) \, m_s \sGs \right] ,
		  \label{eq:OPE1}
\end{eqnarray}
where the function $I_n(M, \sthd)$ is defined by
\begin{eqnarray}
  I_n(M, \sthd) &\equiv& \int_0^{\sth}\!dq_0\> q_0^{n} e^{-\frac{q_0^2}{M^2}} .
\end{eqnarray}
In order to eliminate $|\lambda_\pm|^2$, we differentiate Eq.~(\ref{eq:OPE1}) by $-1/\MB^2$ 
and obtain
  \begin{eqnarray}
  |\lambda_\pm|^2 m_{\pm}^2e^{-\frac{{M_{\pm}}^2}{M^2}}&=&
  \frac{1}{3!4!2^7\pi^6}
  \left[\frac{1}{5600\pi^2}\, I_{13}(M, \sthd)
  \pm \frac{1}{800\pi^2}\, I_{12}(M, \sthd) m_s 
  \right.\nonumber\\
  & &\hspace{3em} 
  \mp\frac{1}{20}\, I_{10}(M, \sthd) \sbs
  + \frac{1}{10}\, I_9(M, \sthd)m_s \sbs 
  \nonumber\\
  & &\hspace{5em} 
  +\frac{1}{40}\, I_9(M, \sthd)\GG
   \pm\frac{1}{4}\, I_8(M, \sthd) \sGs 
   \nonumber \\
  & &\left.\hspace{5em}
		- \frac{1}{4}\, I_7(M, \sthd) \, m_s \sGs \right] .
   \label{eq:OPE2}
 \end{eqnarray}

One sees that the difference between the positive and negative parity states comes from
the terms with $\pm$ sign.  These are the terms which are chirally odd, and thus 
the mass splitting is attributed to the chiral symmetry breaking.  
The leading (i.e., lowest dimension) OPE term which causes the parity splitting is the $m_s$ term, but the contributions of the $\sbs$ and $\sGs$ are larger in magnitude.
At dimension 5, we have neglected $m_s \GG$ term, which happens to be small.  

Dividing Eq.~(\ref{eq:OPE2}) by Eq.~(\ref{eq:OPE1}), we express the masses $m_{\pm}$ 
in terms of the QCD parameters, $m_s$, $\sbs$, $\sGs$, $\GG$, as well as the threshold parameter $\sthd$ and the ``Borel mass'', $\MB$.
The values of the parameters are summarized in Table~\ref{tab:table1}.  Three values of $\sthd$ are chosen for
the evaluation, $\sth = 1.6$, $1.8$, and $2.0$ GeV, while $\MB$ of the range $1.0\to 2.0$ GeV is
considered.
\begin{table}
\caption{\label{tab:table1}
Standard values of the QCD parameters.}
\begin{ruledtabular}
\begin{tabular}{cccc}
$m_s$ & $\sbs$  &  $m_0^2 \equiv \sGs / \sbs$ 
& $\GG$\\
\hline
$0.12$ GeV/c$^2$ & 
$  0.8 \times (-0.23~ {\rm GeV})^3$ & $0.8~{\rm GeV}^2$ & $(0.33~{\rm GeV})^4$  \\
\end{tabular}
\end{ruledtabular}
\end{table}

\begin{figure}[hbtp]
 \caption{
Contributions from each term of Eq.~(\ref{eq:OPE1}) added up subsequently
for the negative-parity and positive-parity sum rules with $\sth=1.8$ GeV.%
}
\begin{center}
\includegraphics*[width=10cm]{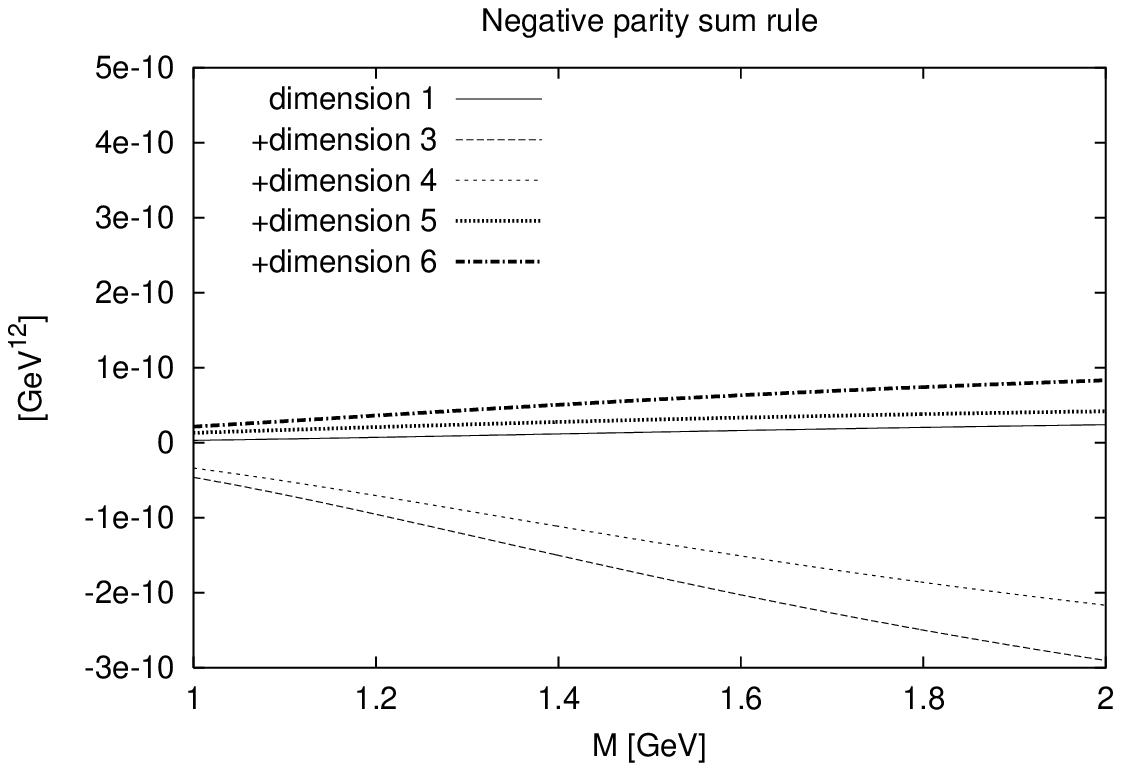}
\includegraphics*[width=10cm]{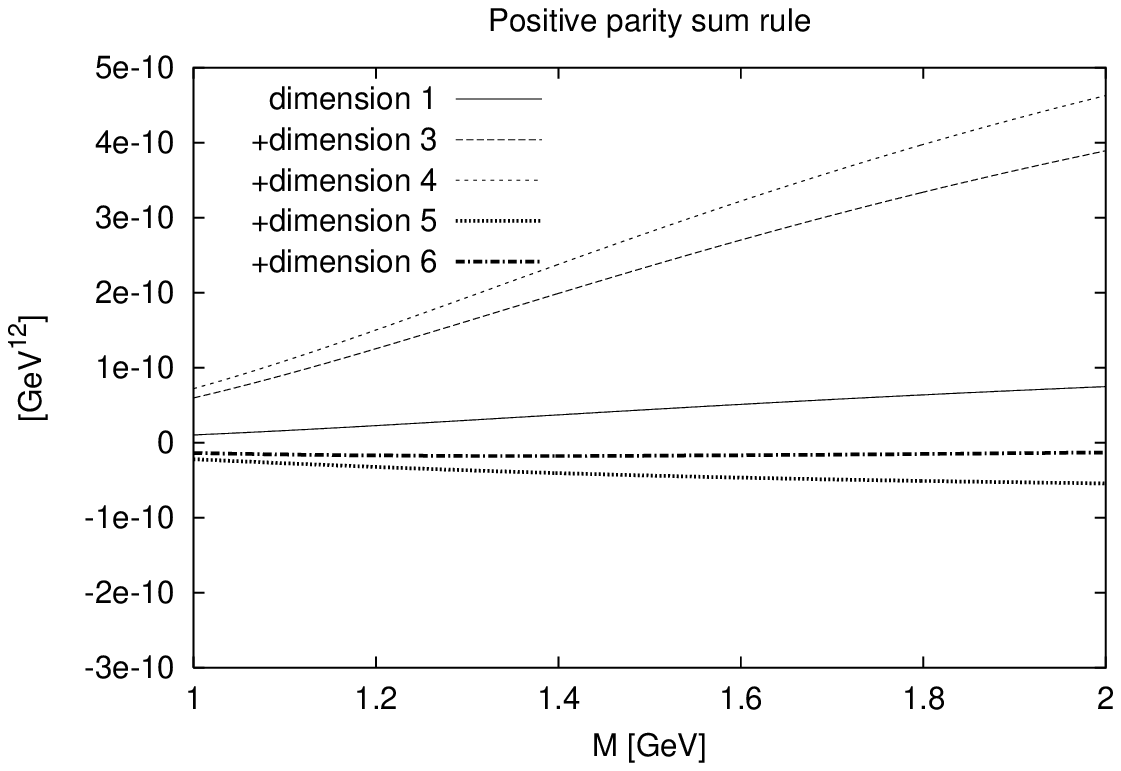}
\end{center}
\label{fig:R}
\end{figure}

The pole residue $|\lambda_{\pm}|^2$ must be positive in order to be able to normalize the baryon state.  In fact, if we find zero or negative residue, such a pole must be spurious.
In order to avoid the spurious pole, we examine the values of Eq.~(\ref{eq:OPE1}), which
are plotted v.s.\  $\MB$ in Fig.~\ref{fig:R}.  There the contributions from each term of OPE are added up subsequently. We find that the dimension-5 condensate, $\sGs$, gives a large negative contribution to 
$|\lambda_{+}|^2$, which ends up with almost zero or a slightly negative value.  
This suggests that the pole in the positive-parity spectral function is spurious.
It is indeed shown that the derived mass for the positive-parity baryon is wildly sensitive to $\MB$ and 
the continuum threshold, $\sth$, as Eq.~(\ref{eq:OPE1}) comes in the denominator of the sum rule.
Therefore we conclude that the sum rule shows no positive-parity solution.

In contrast, the large $\sGs$ contribution makes $|\lambda_{-}|^2$ positive, and therefore
the obtained negative-parity state is a real one.
It is, however, noted that the cancellation between the dimension-3 term, $\sbs$, 
and dimension-5 term, $\sGs$, is rather sensitive to the value of $m_0^2$, defined by
$$m_0^2 \equiv {\sGs \over \sbs}\, .$$
The value is determined from the sum rules of the strange baryons, and a generally accepted value is
$m_0^2 \simeq 0.8 \pm 0.2\, {\rm GeV}^2$.
We therefore vary $m_0^2$ from 0.6 ${\rm GeV}^2$ to 1.0 ${\rm GeV}^2$, and check the positivity 
of $|\lambda_{\pm}|^2$.  
It is found that the conclusion does not change within this window.  
It should be mentioned, however, that if $m_0^2$ were smaller as
0.4 ${\rm GeV}^2$, then the negative parity baryon would be turned to a spurious state, 
while the positive parity becomes a real state.  But  such a value of $m_0^2$ may not be physical.
In Fig.~\ref{fig:R}, one sees that the other terms of OPE are not important, and 
the results are found to be insensitive to the other QCD parameters.
Thus we conclude that the sum rule with the standard values of condensate values predicts 
a negative-parity $\Tp$.

\begin{figure}[hbtp]
 \caption{
Masses of the negative-parity $\Tp$ baryon v.s.\ $\MB$.%
}
\begin{center}
\includegraphics*[width=10cm]{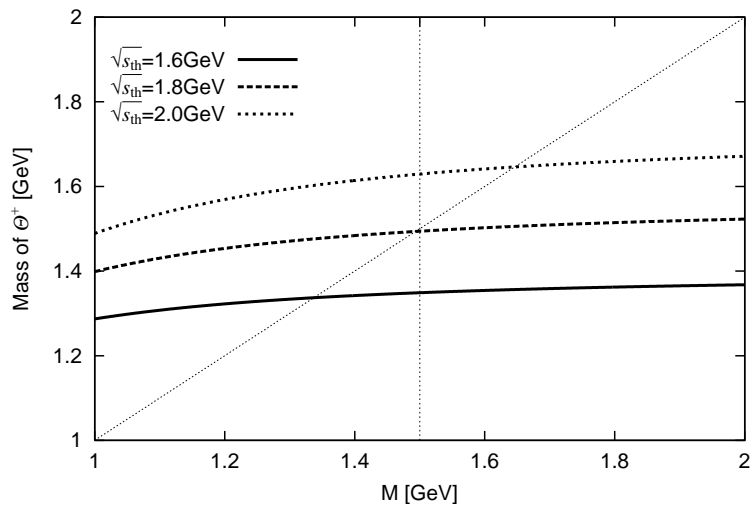}
\end{center}
\label{fig:mass}
\end{figure}

In Fig.~\ref{fig:mass},  the obtained masses of the negative-parity $\Tp$ are plotted 
against $\MB$, where the lines, $\MB=m_{-}$ and $\MB=1.5$ GeV, are also drawn for guidelines. 
The curves show that the $\MB$ dependence is weak and therefore the sum rule works.
The results, however, depend on the choice of the threshold $\sth$ of the continuum,
which may come mainly from the $S$-wave $KN$ scattering states.  
As we expect no excited resonance states in this channel, the continuum starts up gradually,
and therefore the threshold parameter can be as large as 2 GeV.
We thus choose $\sth = 1.6, 1.8$ and 2.0 GeV.
The extracted $\Tp$ masses are given in Table~\ref{tab:tab1}. 
For $\sth=1.8$ GeV, the solutions for $\MB=m_{-}$ and $\MB=1.5$ GeV agree at 
$m_{-} \simeq 1.5$ GeV, which is consistent with the observed $\Tp$ mass.

\begin{table}[hbtp]
\caption{\label{tab:tab1}Masses of the $1/2^-$ baryon for various $\sth$.}
\begin{ruledtabular}
\begin{tabular}{ccc}
$\sth\,[\mathrm{GeV}]$&$M=m_{-}\,[\mathrm{GeV}]$&$M=1.5\mathrm{GeV}\,[\mathrm{GeV}]$\\
\hline
$1.6$&$1.34$&$1.35$\\
$1.7$&$1.42$&$1.42$\\
$1.8$&$1.49$&$1.49$\\
$1.9$&$1.57$&$1.56$\\
$2.0$&$1.65$&$1.63$\\
\end{tabular} 
\end{ruledtabular}
\end{table}

In conclusion, we have performed a QCD sum rule analysis of the penta-quark baryon with strangeness $+1$.
The parity of the state is projected and we have found that the standard QCD condensate parameters
give the negative parity baryon with its mass around 1.5 GeV, although the sum rule is found to be sensitive to the
dimension-5 condensate, $\sGs$, or its ratio to the dimension-3 quark condensate, $m_0^2$.

Finally, we give some comments on previous works on the QCD sum rule approach to the penta-quark baryon.
Zhu\cite{Zhu} performed an analysis of the penta-quark baryon with $I=0, 1$ and 2, in the QCD sum rule.
Matheus et al.\cite{Lee} also calculated the mass of $\Tp$ and $N(1440)$ as penta-quark states in the QCD sum rule.
They both obtain baryon state consistent with the observed $\Tp(1540)$, although the interpolating field operator they used are different with each other, and also from us.
Neither of them, however, determined the parity of the state.  (Although Zhu conjectured the parity is negative, his sum rule cannot give the reason for the conjecture.)
In fact, they both considered only the chiral-even part ($A(q_0)$ part in Eq.~(\ref{eq:AB})) of the correlation function and therefore their results cannot distinguish parity.
On the other hand, we have found that the chiral-odd part ($B(q_0)$ part in Eq.~(\ref{eq:AB})) plays the critical role in determining the parity.
Although the chiral even part of our sum rule gives a consistent result with theirs, we think that the chiral odd
part cannot be ignored.

\bigbreak
\noindent Acknowledgment:
We acknowledge Drs. T.~Nakano, and A.~Hosaka for fruitful discussions.
T.D. acknowledges the support 
of the Japan Society for the Promotion of Science (JSPS).


\def \pnum(#1,#2,#3){{\bf{#1}} 
\ifnum#2>1000 (#2) \else \ifnum#2>53 (19#2) \else (20#2) \fi\fi#3}

\def \NP(#1,#2,#3){{Nucl. Phys.}\ \pnum(#1,#2,#3)}
\def \PL(#1,#2,#3){{Phys.\ Lett.}\ \pnum(#1,#2,#3)}
\def \PRL(#1,#2,#3){{Phys.\ Rev.\ Lett.}\ \pnum(#1,#2,#3)}
\def \PRp(#1,#2,#3){{Phys.\ Rep.}\ \pnum(#1,#2,#3)}
\def \PR(#1,#2,#3){{Phys.\ Rev.}\ \pnum(#1,#2,#3)}
\def \PTP(#1,#2,#3){{Prog.\ Theor.\ Phys.}\ \pnum(#1,#2,#3)}
\def \ibid(#1,#2,#3){{\it ibid.}\ \pnum(#1,#2,#3)}
\def\MO{M.~Oka} \def\etal{{\it et al.}}


\end{document}